\documentclass[12pt,aip]{article}
\begin{document}
\title{On Lyapunov boundary control of unstable magnetohydrodymanic plasmas}
\author{H. Tasso\footnote{het@ipp.mpg.de}, G. N.
Throumoulopoulos\footnote{gthroum@uoi.gr} \\
$^\star$
Max-Planck-Institut f\"{u}r Plasmaphysik \\
Euratom Association \\  85748 Garching bei M\"{u}nchen, Germany \\
$^\dag$ University of Ioannina,\\ Association Euratom-Hellenic
Republic,\\ Department of Physics, GR 451 10 Ioannina, Greece}
\maketitle

\newpage

\begin{abstract}

Starting from a simple marginally stable model considered for Lyapunov based boundary control of flexible mechanical systems, we add a term driving an instability and prove that for an appropriate control condition the system can become Lyapunov stable. A similar  approximate extension  is found for the general energy principle of linearized magnetohydrodynamics. The implementation of such external instantaneous actions may, however, impose challenging constraints for fusion plasmas.  \newline


\end{abstract}

\newpage

\section{Introduction}
Plasmas surrounded by resistive walls may be unstable within free boundary magnetohydrodynamic theory for fusion relevant ${\beta}$ values (See \cite{pt,tt}). To overcome this difficulty, a boundary control or feedback may be necessary (See \cite{js}-\cite{zg}). Lyapunov methods have been applied successfully to the less demanding problem of boundary control of "flexible mechanical systems" as demonstrated in Refs.\cite{qd}-\cite{kr}.

The simplest model of that kind \cite{qd} is governed by the equation of the displacement $u(x,t)$ on the interval $[0,L]$ of a flexible string:
\begin{equation}
u_{tt} - u_{xx} = 0
\end{equation}
with $u(0,t) = 0$, $u_{x}(L,t) = - ku_{t}(L,t)$ with $k > 0$.
To obtain a Lyapunov function for that model, we multiply Eq.(1) by $u_{t}$ and integrate over $x$ from $0$ to $L$ which leads, after integration by parts, to

\begin{equation}
\frac{1}{2}\frac{\partial}{\partial t} \int_{0}^{L}(u_{t}^{2} + u_{x}^{2})dx - [u_{t}u_{x}]_{0}^{L} = 0.
\end{equation}
Applying the boundary conditions above one obtains

\begin{equation}
\frac{1}{2}\frac{\partial}{\partial t} \int_{0}^{L}(u_{t}^{2} + u_{x}^{2})dx  = - ku_{t}(L,t)^{2}.
\end{equation}
The integral is positive and its time derivative is negative, which fulfills the condition for Lyapunov stability. However, the system is marginal for $k = 0$ i.e. without boundary action. Since the relevant plasmas for higher $\beta$ values are not only marginal but rather strongly unstable, the model (1) must be modified to permit an unstable situation in the absence of an external action which is the purpose of section 2.

\section{Unstable model}

The model now considered is a slight modification of Eq.(1) and is given by

\begin{equation}
u_{tt} - u_{xx} - \alpha u = 0
\end{equation}
with $\alpha > 0$, $u(0,t) = 0$, $u_{x}(L,t) = - k_{1}u_{t}(L,t) -k_{2}u(L,t)$ where $k_{1}$ and $k_{2}$ are positive constants. Multiplying Eq.(4) by $u_{t}$ we obtain

\begin{equation}
\frac{1}{2}\frac{\partial}{\partial t} \int_{0}^{L}(u_{t}^{2} + u_{x}^{2} -\alpha u^{2})dx - [u_{t}u_{x}]_{0}^{L} = 0
\end{equation}
after integration by parts over $x$. Inserting the new boundary conditions in (5) we have

\begin{equation}
\frac{1}{2}\frac{\partial}{\partial t} \int_{0}^{L}(u_{t}^{2} + u_{x}^{2} -\alpha u^{2})dx + k_{2}\frac{1}{2}\frac{\partial}{\partial t}u^{2}(L,t) = -k_{1}u_{t}(L,t)^{2}.
\end{equation}

We prove that

\begin{equation}
V = \int_{0}^{L}[u_{t}^{2} + u_{x}^{2} - \alpha u^{2} + \frac{k_{2}}{L}u^{2}(L,t)]dx
\end{equation}
is indefinite for $k_{2} = 0$ if $\pi/4 < L\sqrt{\alpha} < 3 \pi/4$  in the limit $\dot{a}=0$ below, which by the second Lyapunov theorem means that the system is unstable without external action. Indeed, let us first minimize  $\int_{0}^{L}(u_{x}^{2} - \alpha u^{2})dx$ for $u(0,t) = 0$. It leads to $u_{xx} +\alpha u = 0$ whose solution is $u(x,t) = a(t) sin(x\sqrt{\alpha})$. Then inserting the solution in (7)  we obtain
\begin{eqnarray}
V_{min} &=& \int_{0}^{L}\left\lbrack \dot{a}^2sin^{2}(x\sqrt{\alpha}) + a^{2}\alpha cos^{2}(x\sqrt{\alpha}) - a^{2}\alpha sin^{2}(x\sqrt{\alpha}) \right. \nonumber \\ & & \left. + \frac{k_{2}}{L}a^{2}sin^{2}(L\sqrt{\alpha})\right\rbrack dx
\end{eqnarray}
or
\begin{eqnarray}
V_{min}& =& \int_{0}^{L}\left\lbrack \dot{a}^2sin^{2}(x\sqrt{\alpha})
+ a^{2} \alpha cos(2x\sqrt{\alpha}) \right.  \nonumber \\ 
& & \left.   + \frac{k_{2}}{L}a^{2}sin^{2}(L\sqrt{\alpha})\right\rbrack dx.
\end{eqnarray}
Finally
\begin{equation}
V_{min} =\frac{\dot{a}^2}{2}\left(L- \frac{sin(2L\sqrt{\alpha})}{2 \sqrt{\alpha}}\right) + \frac{a^2 \sqrt{\alpha}}{2} sin(2L\sqrt{\alpha}) +  k_{2}a^2 sin^{2}(L\sqrt{\alpha}).
\end{equation}
The system can be stabilized if
\begin{equation}
k_{2} > -\left\lbrack  \frac{\dot{a}^2}{2}\left(L- \frac{sin(2L\sqrt{\alpha})}{2 \sqrt{\alpha}}\right) + \frac{a^2 \sqrt{\alpha}}{2} sin(2L\sqrt{\alpha}) \right\rbrack \left( a^2 sin^{2}(L\sqrt{\alpha}) \right)^{-1},
\end{equation}
which in the limit of $\dot{a}=0$ becomes
\begin{equation}
k_{2} > -\frac{\sqrt{\alpha}}{2}\frac{sin(2L\sqrt{\alpha})}{sin^{2}(L\sqrt{\alpha})}.
\end{equation}

\section{Magnetohydrodynamic boundary control}
From the models above we have learned that $k_{2}$ is essential to influence positively the Lyapunov function. Let us now try to apply the procedure to the energy principle of linearized magnetohydrodynamics (See \cite{pt,tt,bf})
\begin{equation}
\frac{1}{2}\frac{\partial}{\partial t}[(\dot{\bf\xi}.\dot{\bf\xi}) + \delta W_{P} + \delta W_{P,W} + \delta W_{W,inf}] = -\sigma\int_{W}(\dot{A})^{2}dS
\end{equation}
where ${\bf\xi}$ is the plasma displacement and the subscripts refer to the plasma, vacuum between plasma and wall and  vacuum between wall and infinity respectively. At the plasma vacuum interface, we have ${\bf n}\times {\bf A} = ({\bf n}\cdot{\bf\xi}){\bf B}_0 $ where ${\bf n}$ is the normal to the interface, ${\bf A}$ is the perturbed vector potential at the vacuum side and ${\bf B}_0$ is the equilibrium magnetic field. $\sigma$ is the conductivity of the wall. The last volume integral in (13) can be transformed in a surface integral
\begin{equation}
\frac{1}{2}\frac{\partial}{\partial t}[\delta W_{W,inf}] =
\frac{1}{2}\frac{\partial}{\partial t}\int_{W,inf} (\nabla\times{\bf A})^{2}d\tau = \frac{1}{2}\int_S {\bf n}_S\cdot{\bf \dot{\bf A}}\times \nabla\times{\bf A} dS
\end{equation}
where $S$ is the external surface of the wall and ${\bf n}_{S}$ is the normal to that surface. Eq.(13) can now be written
\begin{equation}
\frac{1}{2}\frac{\partial}{\partial t}[(\dot{\bf\xi}.\dot{\bf\xi}) + \delta W_{P} + \delta W_{P,W} + \int_S {\bf n}_S\cdot{\bf \dot{\bf A}}\times \nabla\times{\bf A}dS] = -\sigma\int_{W}(\dot{A})^{2}dS
\end{equation}
At this point it is usually assumed that ${\bf n}_S\times{\bf A} = 0$ if the wall is taken as perfectly conducting. An alternative is to choose a boundary control law like
\begin{equation}
{\bf n}_S\times\nabla\times{\bf A} \rightarrow {\bf n}_S\times\nabla\times{\bf A} -k_{2} {\bf A}  
\end{equation}
on $S$ with $k_{2}$ constant in time but as a free function within $S$. Then (15) becomes
\begin{equation}
\frac{1}{2}\frac{\partial}{\partial t}[(\dot{\bf\xi}.\dot{\bf\xi}) + \delta W_{P} + \delta W_{P,W} + \delta W_{W,inf} + \int_S k_{2}{\bf A}^{2}dS] = -\sigma\int_{W}(\dot{A})^{2}dS
\end{equation}
Similarly to section 2 we can minimize, for $\dot{\xi} = 0$, $\delta W = \delta W_{P} + \delta W_{P,W} + \delta W_{W,inf}$ with respect to $({\bf\xi} , {\bf A})$ with ${\bf A}$ vanishing for large distances to obtain $({\bf\xi}_{min} , {\bf A}_{min})$ and $\delta W_{min}$.
Since $\delta W_{min}$ is indefinite the condition for stabilization is
\begin{equation}
\int_S k_{2}{\bf A}_{min}^{2} \geq \delta W_{min}.
\end{equation}
We can now determine $k_{2,min}$, by optimizing the solutions of (18), as a function of the position on the surface or, simply, assume  a positive constant $k_{2}$ satisfying (18).

Let us notice that in the boundary law (16) a magnetic field is imposed on the surface S, which implies a surface current density. This is possible only if that surface is perfectly conducting. 
For resistive conductors the term $\int_S k_{2}{\bf A}^{2}dS$ will disappear in (17). Strictly speaking, a  surface boundary control is not possible in the magnetohydrodynamic case unlike the case of section 2. In real situation  the control on a boundary layer of (small) finite  thickness seems necessary and such a treatment requires  further investigation.

\section{Discussion}

An unstable model, in contrast to the marginal models of Ref.\cite{qd}, can be stabilized by controlling the boundary  if the external action (here $k_{2}$) is large enough. It turns out that the method used for this model can be extended to linearized magnetohydrodynamics for which codes exist to minimize the energy principle of Ref.\cite{bf}  in the approximation that the external surface of the wall is covered by an infinitesimally thick layer of perfectly conducting actuators. In addition to potentially extending the study by removing this assumption many problems remain open of how to realize instantaneous boundary conditions by coils in view of their induction, resistance etc. Starting from the sensors, the amplifyers should have high saturation levels and broad bands. Moreover,
since the actuators must consist of coils, it means that they may have to be placed within the resistive vacuum vessel to avoid time delays. This may be a challenging problem for fusion reactors.

\begin{center}

{\large\bf Acknowledgements}

\end{center}

Part of this work was conducted during a visit of one of the
authors (G.N.T.) to the Max-Planck-Institut f\"{u}r Plasmaphysik,
Garching. The hospitality of that Institute is greatly
appreciated. This work was performed within the participation of the University of Ioannina in the Association Euratom-Hellenic Republic, which is supported in part by the European Union and by the General Secretariat of Research and Technology of Greece.  The views and opinions
expressed herein do not necessarily reflect those of the European
Commission.


\newpage





\vspace*{-2cm}

\begin{thebibliography}{99}



\bibitem{pt} D. Pfirsch, H. Tasso, Nuclear Fusion 11, 259 (1971).

\bibitem{tt} H. Tasso, G.N. Throumoulopoulos, Physics of Plasmas 18, 070702 (2011).

\bibitem{js} S.C. Jardin, J.A. Schmidt, Nucl. Fusion 38, 1105 (1998).

\bibitem{gj} A.M. Garofalo, T.H. Jensen, E.J. Strait, Phys. Plasmas 9, 4573 (2002).

\bibitem{bo} A.H. Boozer, Phys. Plasmas 11, 110 (2004). 

\bibitem{mp} S.Yu. Medvedev and V.D. Pustovitov,  Plasma Phys. Reports 30, 895 (2004).
\bibitem{by}P.R. Brunsell et al., Phys. Plasmas 12, 092508 (2005).

\bibitem{sw} E. Schuster, et al. Automatica 41, 1173 (2005).

\bibitem{sb} S.A. Sabbagh et al., PRL 97, 045004 (2006).

\bibitem{li} Y. Liu,  Plasma Phys. Control. Fusion 51, 115006 (2009).

\bibitem{fi} R. Fitzpatrick, Nucl. Fusion 51, 053007 (2011).

\bibitem{vg} F Villone et al., Plasma Phys. Control. Fusion 54, 085003 (2012).

\bibitem{il} V. Igochine, Nucl. Fusion 52, 074010 (2012).

\bibitem{zg} L.E. Zakharov, S.A. Galkin, S.N. Gerasimov, Phys. Plasmas 19, 055703 (2012).


\bibitem{qd} M.S. de Queiroz, D.M. Dawson, S.P. Nagarkatti, F. Zhang, "Lyapunov-based Control of Mechanical Systems", Birkh\"{a}euser ISBN 978-0-8176-4086-6 (2000).  

\bibitem{kr}  M. Krstic, A. Smyshlyaev,  R. Vazquez et al., {\it Boundary control of PDEs: A short course on backstepping designs}, http://flyingv.ucsd.edu/pde.pdf.

\bibitem{kr}  M. Krstic,  International Journal of Robust and Nonlinear Control. 16,  801  (2006).


\bibitem{bf} I. Bernstein, E.A. Frieman, M.D. Kruskal, R.M. Kulsrud, Proc. Roy. Soc. (London) A 244, 17 (1958). 




\end{thebibliography}
\end{document}